*Original Article*

# Cyber-Resilient Privacy Preservation and Secure Billing Approach for Smart Energy Metering Devices

M. Venkatesh Kumar[1], C. Lakshmi[2]

[1,2]Department of CSE, SRM Institute of Science and Technology, Chennai, Tamilnadu. India.

[1]Corresponding Author : veenkat@gmail.com

***Abstract*** - *Most of the smart applications, such as smart energy metering devices, demand strong privacy preservation to strengthen data privacy. However, it is difficult to protect the privacy of the smart device data, especially on the client side. It is mainly because payment for billing is computed by the server deployed at the client's side, and it is highly challenging to prevent the leakage of client's information to unauthorised users. Various researchers have discussed this problem and have proposed different privacy preservation techniques. Conventional techniques suffer from the problem of high computational and communication overload on the client side. In addition, the performance of these techniques deteriorates due to computational complexity and their inability to handle the security of large-scale data. Due to these limitations, it becomes easy for the attackers to introduce malicious attacks on the server, posing a significant threat to data security. In this context, this proposal intends to design novel privacy preservation and secure billing framework using deep learning techniques to ensure data security in smart energy metering devices. This research aims to overcome the limitations of the existing techniques to achieve robust privacy preservation in smart devices and increase the cyber resilience of these devices.*

## 1. Introduction

Most of the real-time smart applications such as data sharing, smart grid systems, smart energy devices etc. can be simulated using a fundamental client-server structure where the client connects with the server to send requests and receive requested services from the services (De Craemer&Deconinck, 2010) [1]. This architecture also allows the clients to pay for the received services. For instance, smart devices allow users to contact the centralised server through their smart gadgets to obtain relevant knowledge. Applications like smart metering also follow a basic client-server system wherein smart meters are deployed on the user's side to collect the electricity usage data (Arun & Mohit, 2016) [2]. This data will be further communicated to the electrical companies via servers (deployed on the company's side). This process of data sharing is more vulnerable to cyber attacks since the data sent from the client to the server can expose sensitive user data, such as the user's private information, to unauthorised entities (Cheng et al., 2018) [3]. The attackers can exploit sensitive information such as the user's identity and financial data to find out the preferences of the users (Chim et al., 2012) [4] (Li et al., 2015) [5] (Xing et al., 2017) [6]. Readings from smart meters may be analysed to determine whether or not a user is there, and this information can be dangerous. (Garcia &Jacobs, 2010) [7] Since they can trigger security and safety concerns of the users. In such cases, privacy preservation becomes a crucial tool in smart metering devices. In addition, the payment made by the client (electricity bill) to the energy companies via a server is also exposed to cyber-attacks. Hence it is essential to maintain the client's privacy and secure the billing information using privacy preservation techniques. Conventional cryptographic techniques were used in various research works for privacy preservation in smart energy metering devices (Li et al., 2014) [8] (Yao et al., 2019) [9] (Syed et al., 2020) [10]. However, these techniques suffer from certain problems which restrict their adaptability. These techniques suffer from high computational overhead and are not feasible for smart applications (Paulet et al., 2013) [11] (Huang et al., 2014) [12] (Chen et al., 2019) [13]. Also, it is highly difficult to hide clients' information from unauthorised users when they communicate through the servers since the requests and payments made by the clients follow a traditional billing process.

In this process, the server computes the bills based on the utilised service, and the information about the payment made by the client is easily accessible. It increases security problems (Molina-Markham et al., 2012) [14] (Alhothaily et al., 2017) [15]. Most of the existing privacy preservation techniques for smart metering (Erkin et al., 2013) [16] (Shen et al., 2017) [36] (Wang, 2017) [18] could not meet the desired requirements for privacy preservation since they compromise on the service quality for privacy preservation.

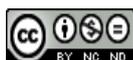

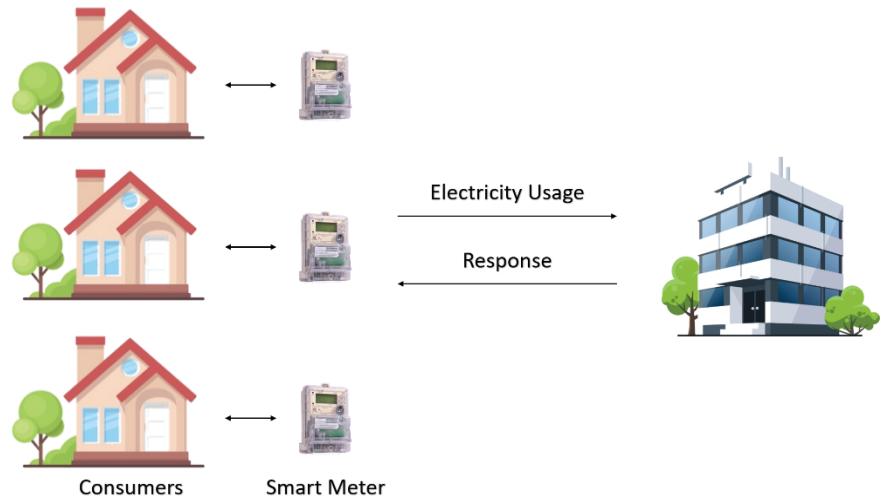

**Fig. 1 Smart Metering Architecture**

In this context, there is a great demand for cyber-resilient systems that protect smart devices from adversarial cyber-attacks. In general, cyber-resilient systems are the systems which can predict, withstand and recover from adverse cyber-attacks. Since smart energy metering devices are more susceptible to cyber-attacks, making these systems cyber-resilient is essential to protect them from being exploited by attackers. With the advancement of machine learning and deep learning techniques, they are used in smart metering devices for privacy preservation. This research proposes a novel deep learning-based privacy preservation approach which is cyber-resilient for secure billing in smart energy metering devices. The research methodology section discusses a brief description of the proposed approach.

The Organization of the paper is as follows: Section 2 Security vs Privacy, Section 3 Cyber resilience and Privacy Preservation, Section 4 Literature survey, Section 5 Problem statement, Section 6 Aim and objective, Section 7 Proposed research methodology, Section 8 Experimental results, Section 9 Discussion and future research and Section 10 Conclusion.

## 2. Security Vs Privacy

Privacy and security are prioritised, protecting data against misuse and leakage during the data mining process. The main difference is security deals with safeguarding, and privacy deals with safeguarding the user's identity.

| Security | Privacy |
|---|---|
| Confidentiality, availability, and integrity are referred to as data security. | Related to the appropriate use of data. |
| Controlling of unauthorised or modification of information during the mining process. | Preventing or not disclosing individual or group-related sensitive information. |

## 3. Cyber Resilience and Privacy Preservation

Cyber resilience is defined as the ability of an organisation to enable business acceleration by preparing for, responding to and recovering from threats, i.e. cyber threats. Cyber resilience is very important because traditional security is not enough to ensure adequate security like information security, data security and network security. The difference between Cyber security and Cyber resilience is as follows:

| Cyber Security | Cyber resilience |
|---|---|
| Refers to the methods and processes of protecting electronic data, which includes identifying data, technology, location and protection. | Refers to withstanding and recovering from threats which disrupt business operations. |

### 3.1. Privacy Preservation
The different models for privacy-preserving were (i). Trust third-party model, (ii). Semi-honest model, (iii). Malicious model, and (iv). Other models. In the third-party trust model, the third party performs the computation. It delivers the results by following secure protocols, whereas, in a semi-honest model, inputs were used according to the protocols. There are no protocols or restrictions on the participants in the malicious model; in other models, cryptographic techniques were carried out for performing data mining tasks.

## 4. Literature Survey

Privacy preservation in smart metering devices has gained huge attention among researchers in recent times (Rial &Danezis, 2011) [19] (Souri et al., 2014) [20] (Mustafa et al., 2015) [21] (Gohar et al., 2019) [37]. The emergence of deep learning techniques and privacy preservation in smart metering has witnessed a significant transformation. Various researchers have proposed deep learning-based techniques for protecting sensitive data and users' privacy from cyber-attacks. (Joudaki et al., 2020)

[23] Proposed a comprehensive analysis of different deep learning algorithms for privacy preservation in smart grids. The efficacy of deep learning algorithms in maintaining the security of smart grids is stated in the review. Deep learning algorithms are also effective in identifying intrusions and energy theft. Different algorithms, such as neural networks, reinforcement learning algorithms, and recurrent neural networks, can strengthen privacy in smart grid-based applications. However, this review focuses mainly on supervised learning techniques, and more research is required to investigate the potential ability of unsupervised learning algorithms to protect privacy in smart grids. In a survey conducted by (Asghar et al., 2017) [24], it was stated that there is a lack of focus. The study looked at power and security distributed machine learning techniques for privacy protection...The investigation focused on the security of smart meter data utilising various cryptographic techniques, as well as other privacy-preserving data management systems that did not need the usage of a non-trusted operating system. A similar review was conducted by (Giaconi et al., 2020) [25], wherein the review focused on the deployment of smart metering for smart grid applications. The study focused on analysing machine learning and big data techniques for improving privacy preservation in smart metering. The study highlights the prominent advantages and discusses the challenges associated with smart metering. It can be inferred from existing studies that both big data and machine learning have a high significance in enhancing the security of user data in smart metering devices. Their adaptation should be extensively explored for more practical applications. (Ibrahem et al., 2020) [26] Proposed a novel and robust privacy-preserving model for Using a dynamic billing and load monitoring strategy to detect power theft. This study mainly focused on the encryption of smart meter data by incorporating a functional encryption approach and enabling the system operator to use cipher texts for computing the electricity bill using a dynamic pricing approach. Both the grid and load functioning were monitored, and a machine learning algorithm was used to identify fraudulent clients without exploiting the readings of the smart meters to protect the user's privacy. In addition, a functional encryption technique was employed for aggregating the encrypted smart meter readings for smart billing and load monitoring so that only the aggregated value is provided to the system operator. Similar to this approach, (Ibrahem et al., 2020) [27] proposed implementing AMI's uses, a deep learning architecture for optimal privacy protection networks by using a secure data collection approach. In the proposed work, a novel technique known as STDL is proposed for secure data collection from smart meters in AMI networks without exploiting consumer data privacy by sending redundant real-time readings using a deep learning algorithm. Initially, a clustering approach is used to aggregate the real-time meter reading for creating a dataset for analysing the data patterns. The attacker model was then trained using a deep learning technique. The experimental results revealed that the suggested approach provides exceptional accuracy of 91% in identifying fraudulent customers and securing consumer data privacy.

## 5. Problem Statement

Privacy preservation is one of the important topics in smart applications such as smart metering and smart billing. More and more smart applications depend on the huge volume of data in the distributed storage collected over time. It becomes difficult to maintain the security and privacy of the data. It is to be noted that most smart applications collect, store and use different types of data; consequently, sharing the data among different entities (clients) is challenging since it imposes a severe threat to data privacy. Hence, most applications depend on privacy-preserving methods for enhancing the security and reliability of the data-sharing process between the client and the server. The majority of real-time applications are built on a client-server design, which allows clients to make requests to the server and pay the server for the services obtained. However, securing clients' privacy (identity and data) while simultaneously obtaining services from the server and paying the server for the service gained within a billing period is extremely complex and demanding under this architecture. There are various challenges in this aspect which need to be addressed. Some of the prominent research problems identified in this research are:

1. It is difficult to maintain the privacy of the client's identity and client-related information while verifying the authenticity of the information.
2. Since the server receives a huge volume of requests from clients, it is challenging to protect the confidentiality of each request.
3. In the billing process, the server sends the monthly bill to the client, and in such cases, there are chances of information leakage by malicious attackers. This problem must be resolved to enhance the security and integrity of the server.
4. The attackers in smart applications such as smart metering link the transactions carried out for multiple billing periods with unauthorised entities. It allows attackers to gain illegal access to the server data and pose a significant threat to data privacy.
5. Existing privacy-preserving techniques suffer from the problem of high computational and communication overload on the client side. It will reduce the performance of these techniques and might lead to inaccurate results.
6. Most of the techniques are not efficient in terms of communication complexity. Most existing privacy preservation solutions depend on a synchronous network sensitive to dynamic adversaries. This decreases the potentiality of these solutions, which do not comply with the security features of the smart applications.

The problems mentioned above must be incorporated while designing effective privacy preservation techniques. In this context, this research intends to develop a potentially robust and novel accountable approach to provide seclusion preservation and bill in client-server-based applications. The approach must be designed in such a way that it can achieve better non-repudiation and accountability so that the clients should not be able to cancel the service requests which have been provided to the server. The server must not be able to counterfeit or replicate the service request without verification or authorisation.

# 6. Aim and Objectives

The proposed research's primary goal is to develop novel privacy preservation and secure billing approach for smart energy metering devices. The research objectives of this work are listed in the points below.

*6.1 Research Objectives*
1. To study and explore deep learning techniques for privacy preservation in smart applications.
2. To develop robust privacy preservation and smart billing framework using deep learning for smart energy metering devices.
3. To strengthen the data security and privacy of the smart energy metering devices by employing a secure multiparty computation approach and data encryption mechanisms.
4. To overcome the issues of computational overhead, communication overhead and dependency on the third party for data sharing.
5. To compare the performance is to validate the efficiency of the suggested technique; it was compared to the existing approach.

# 7. Proposed Research Methodology

This work proposes a novel deep learning-based privacy-preserving and smart billing framework for smart metering devices. The suggested method's major goal is to overcome the problem of computational complexity overhead in large-scale data processing applications such as smart energy metering without compromising on data privacy protection. The study also aims to strengthen the privacy-preserving model against dynamic adversaries and interference threats. Privacy preservation techniques without a trusted third party depend on secure MPC (Atallah &Blanton, 2009) [28]. The novelty of the proposed work is that this study aims to look at the characteristics of secure multiparty computation. (MPC) for privacy preservation without involving any third-party entity. It will reduce the dependency on the third party and strengthen the privacy preservation process. In this research, Multiparty computation will allow a group of smart devices to compute collaboratively without revealing any important information other than required data or relevant data (Volgushev et al., 2019) [29]. Existing privacy preservation techniques assume that the data collected from the dataset is clean and pre-processed with predefined features. This reduces the adaptability of these techniques for advanced and real-time applications. To overcome this problem, this study employs a feature selection approach wherein the relevant features are selected for performing privacy preservation and smart billing.

*7.1 Overview of the proposed approach*
This study proposes a system based on deep learning Long Short Term Memory (Bidirectional) network-based along with a multi-protocol SPDZ (MP-SPDZ) for privacy preservation. The MP-SPDZ is a secure MPC approach which is used in this research to enhance privacy preservation in smart meters. Multi-Protocol SPDZ (MP-SPDZ) is an effective protocol used for the implementation of MPC for privacy preservation in distributed data models (Sharma et al., 2019) [30] (Schoppmann et al., 2018 [31]. The protocols commonly used for privacy preservation comprise homomorphic encryption, oblivious transfer and secret sharing. The MP-SPDZ protocol, when combined with a deep learning model, reduces the computational cost and improves the performance and accuracy of the privacy-preserving models (Dalskov et al., 2020) [32].
On the other hand, Bidirectional LSTM (BI-LSTM) is an advanced version of conventional LSTM models. LSTM is a recurrent neural network used widely in attack detection and classification tasks. Bi-LSTM can train two LSTMs instead of one LSTM on the input sequence and hence is more advantageous compared to the single LSTM model. Bi-LSTM operates in both the forward and backward directions. In every step, it records the previous recordings of the smart meters stored in the memory and evaluates the probability of the next billing cycle. The Bi-LSTM model is a four-layer neural network, and its memory unit, each LSTM, there are three gates in this circuit: an input gate, an output gate, and a forget gate. The model might keep or forget the data at any time by altering the data flow via these components. As a result, the Bi-LSTM model can only keep track of important data.

*7.2 Secure MPC and Deep Learning for Privacy Preservation*
The real-time implementation of the MPC will be achieved by employing an integrated approach of SPDZ protocol combined with a Bidirectional LSTM to perform privacy preservation and smart billing in smart energy metering devices. In this research, the MP-SPDZ protocol will ensure that the data collected from the smart meter is secured before being processed by the server. This protocol will enable the modification of the request or the payment by allowing them to create or delete the data at any time. However, the service providers can monitor and verify the changes made. As part of secure billing, the MP-SPDZ protocol will validate the user credentials, compare them with the available encryptions stored in the database, and execute the instructions after analysing the data. The results are secured in such a way that the original data stored on the server is not accessible to anyone. The data is processed by MP-SPDZ protocol using two stages, i.e., offline mode and online mode. In offline mode, the client's requests are shared with other participating members. In online mode, the actual deep learning task will be performed. The Bidirectional LSTM algorithm will be used to identify the fraudulent customers and the cyber-attacks using the feature selection approach. The sensitive user data will be protected from exploitation by identifying the cyber-attacks. Hence, it will be ensured that the multi-party authorities can access the data only, which is relevant to them, and other important data is preserved.

### 7.3. Workflow
The workflow of the proposed approach is discussed in the below steps:

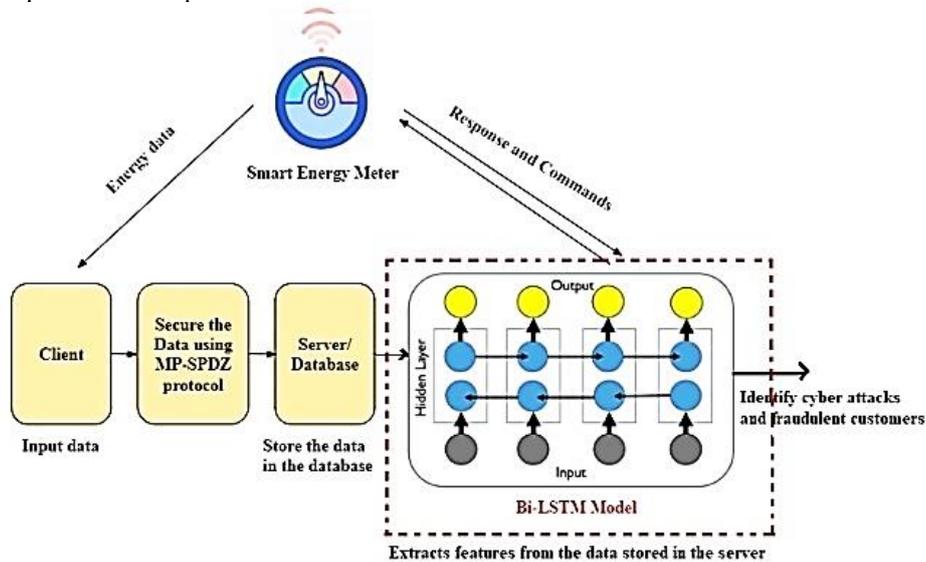

**Fig. 2 Proposed Block Diagram**

#### 7.3.1. Step 1: Identification of Data Sets:
At this stage, the data from the datasets will be identified. Based on the obtained data, privacy-preserving tasks will be performed. Real smart meter data from the Irish Smart Energy Trials [33] can be used for the experimental investigation. Electric Ireland and the Sustainable Energy Authority of Ireland released the statistics in January 2012. Consumers in the dataset submit their measurements every 30 minutes.

On the other hand, another dataset, namely the SustDataED2 dataset (Pereira et al., 2022) [34], can be used. The dataset covers 96 days of utility usage from a single Portuguese home, both averaged and individually. The dataset also contains the timestamps of the observed appliances' ON-OFF transition for the duration of the deployment, which serves as the necessary ground truth for analysing machine learning difficulties.

#### 7.3.2. Step 2: Data Pre-Processing
The raw data collected from the dataset consists of different types of uncertainties. It is essential to process the raw data in order to make it appropriate for data mining tasks. In this stage, the data aggregated from the datasets are subjected to pre-processing.

#### 7.3.3. Step 3: Feature Selection for Privacy Preservation
The study uses a Bi-LSTM to extract relevant features from the textual data and to select appropriate features for privacy preservation. The selected features will be used for detecting fraudulent customers and cyber-attacks.

*Syntax*
*layer = bilstmLayer(numHiddenUnits)*
*Layer = bilstmLayer (numHiddenUnits, Name, Value)*

*Description*
`layer = bilstmLayer(numHiddenUnits)`
*The above statement creates a Bi-LSTM layer and sets the NumHiddenUnits property.*

*Properties*
- NumHiddenUnits
- OutputMode
- HasStateInputs
- HasStateOutputs
- InputSize

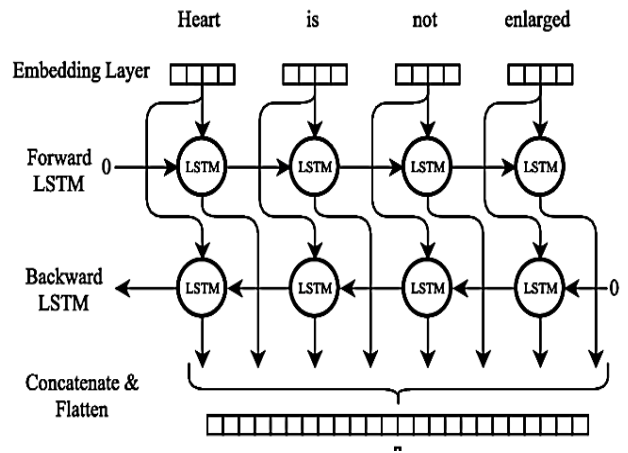

**Fig. 3 Bi-LSTM model**

### 7.3.4. Step 4: Privacy Preservation using Bi-LSTM and MP-SPDZ approach

Based on the extracted features, the Bi-LSTM model will classify the data as normal or malicious and identify fraudulent customers. The MP-SPDZ protocol will ensure the data's security by protecting the data's privacy before injecting it into the server.

### 7.3.5 Step 5: Security analysis

The security properties like client privacy preservation and Authentication were analysed.

*Client Privacy Preservation*

In the proposed Bi-LSTM during the billing term, the client's identity and request will be protected as well as during the multiple billing periods. The Bi-LSTM method is used for detecting the data, i.e. finding out which is normal, fraudulent customers and any cyber-attacks.

*Authentication*

The MP-SPDZ protocol will ensure the security of the data by protecting the privacy of the data before injecting into the server.

### 7.3.6. Step 6: Performance Evaluation

The suggested approach's performance will be measured using several performance measures such as accuracy, precision, recall and memory. In addition, the performance will also be measured for computational overhead and communication overhead. A comparative analysis will be conducted to validate the proposed method's efficacy.

$$\text{Accuracy} = \frac{\text{True Positive}}{(\text{True Positive} + \text{True Negative}) * 100} \quad (1)$$

$$\text{Precision} = \frac{\text{True Positive}}{(\text{True Positive} + \text{False Positive})} \quad (2)$$

$$\text{Recall} = \frac{\text{True Positive}}{(\text{True Positive} + \text{False Negative})} \quad (3)$$

## 8. Experimental Results

The experiment was conducted on INTEL ® CORE™ i5-7300 HQ CPU @ 2.5 GHz 2.50 GHz 8gb, 64bit Windows 10 Operating System X64 Based processor. Programming Languages and experimental tools were python version 3.7.4, Visual Studio code (IDE), Jupyter Notebook (web-based interactive computing platform) and Google Colab.

**Table 1. Data Set Description**

| S.No. | Dataset Name | Type |
|---|---|---|
| 1 | Irish Smart Energy Trials | Multivariate, Time Series |
| 2 | SustDataED2 | Multivariate, Time series |

**Table 2. Performance Evaluation**

| S.No. | Dataset | Accuracy (%) | Memory (%) |
|---|---|---|---|
| 1 | Irish Smart Energy Trials | 92.5 | 85 |
| 2 | SustDataED2 | 91.8 | 89 |

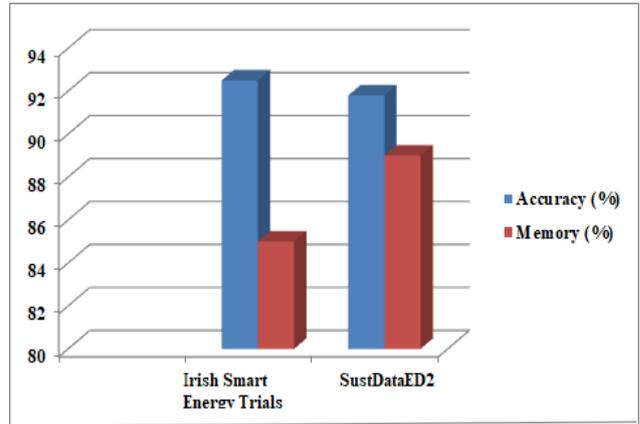

**Fig. 4 Accuracy and Memory**

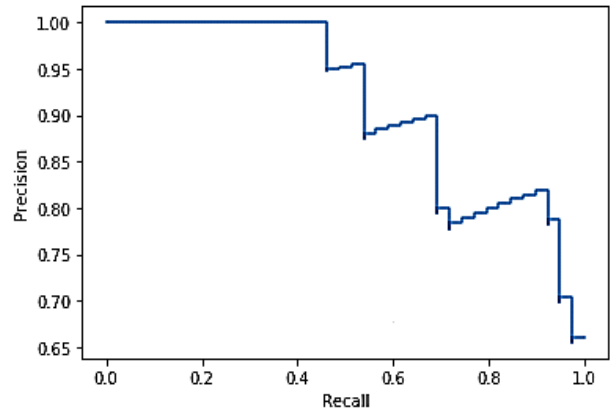

**Fig. 5 Precision and Recall**

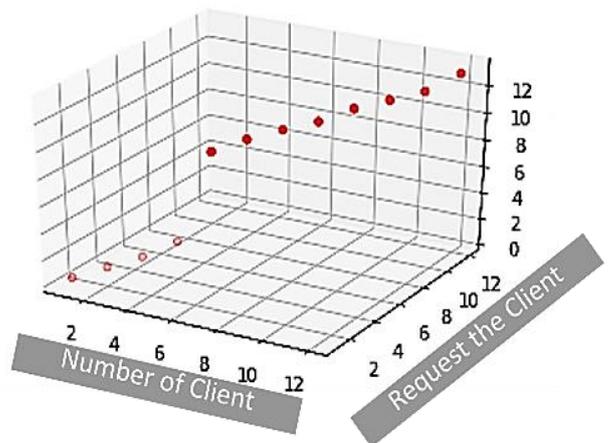

**Fig. 6 Communication Overhead of group of clients in a billing period**

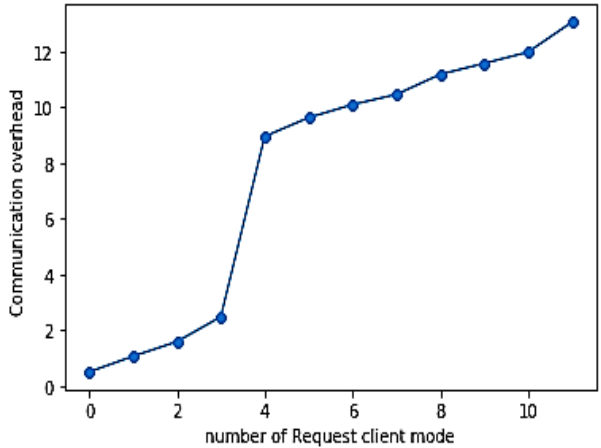

Fig.7 Communication Overhead

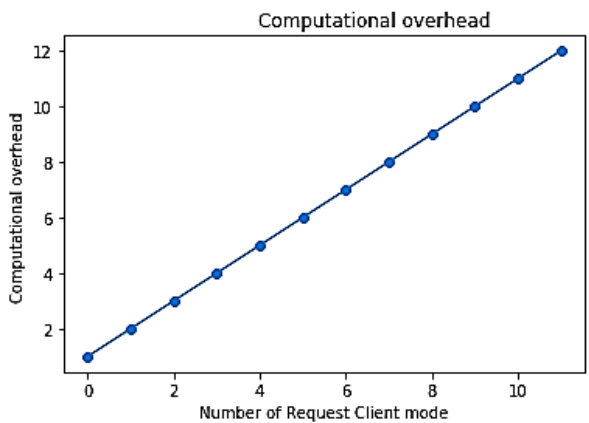

Fig. 8 Computational Overhead

## 9. Discussions and Future Research

Table 1 represents the data set used for the experiment, and table 2 shows the performance evaluation, which shows good accuracy achieved for both datasets, as shown in Fig 4. Fig 5 shows the result relevancy used in the information retrieval, and the relevant information returned. It also shows the false positive rate, i.e. high precision and low false negative rate, i.e. high recall. Fig. 6, 7 and 8 show the communication overhead of a group of clients in a billing period with reduced communication overhead and computational overhead. The proposed approach is used to preserve the data privacy and data security of smart energy metering devices. Also, the secure MPC-based privacy preservation framework is designed to satisfy all the objectives with good performance efficiency in terms of accuracy, security, privacy, reduced computation overhead and communication overhead.

### 9.1. Future Research
In future research, the following two directions can be implemented. First, intend to instantiate the EBCC scheme for privacy-preserving data collection. Secondly, plan to consider the properties of secure multiparty computation. The EBCC can be used for lightweight operations in encryption, aggregation, and decryption which result or improve in low computation and communication overheads. Using this approach, the security analysis would demonstrate that the EBCC will be secured, can resist collusion attacks and hide customers' distribution which is needed for a fair balance checking in credit card payment.

## 10. Conclusion

This research proposes a novel privacy preservation and secure billing approach using a hybrid MP-SPDZ and Bi-LSTM model. The study addresses the problems related to the security and privacy of the client and smart meter data. The proposed approach has addressed the problem of high computational and communication overload on the client side and dependency on the third party for data sharing. The model is designed to operate in two stages; offline and online. Offline mode is related to data sharing between the clients. In the online model, the Bi-LSTM algorithm is trained to identify cyber-attacks using the feature extraction method and secure the privacy of the energy data.